# Enrichissement des contenus par la réindexation des usagers : *un état de l'art sur la problématique*

*Azza HARBAOUI*
Doctorante
ESCE & Université la Manouba
Tunisie

Azza. harbaoui@gmail.com

*Malek GHENIMA*
MA. & Chercheur
ESCE & Université la Manouba
Tunisie

Malek.Ghenima@escem.rnu.tn

*Sahbi SIDHOM*
MCF. & Chercheur
LORIA–Nancy2 & Nancy Université
France

Sahbi.Sidhom@loria.fr

**Résumé :**
La recherche d'information (RI) est une démarche faite par un utilisateur pour obtenir de l'information pertinente qui répond à ses besoins moyennant un système de RI (SRI). Toutefois, les SRI montrent certains écarts entre la pertinence de l'utilisateur et la pertinence du système. Ces écarts sont liés essentiellement à l'imperfection de l'indexation (approche directement liée au processus de RI), aux problèmes de l'incompréhension du langage naturel et à la non correspondance entre le besoin réel de l'utilisateur et la réponse à sa requête. D'où l'idée de penser à une indexation qui prend en compte le point de vue de l'usager. Ce dernier en consultant le document peut construire des informations à valeur ajoutée sur le contenu existant : il s'agit de nouvelles informations qui enrichissent le contenu, qui permettent la visibilité sémantique ou qui facilitent la lecture par les annotations, par des liens vers d'autres contenus, par de nouveaux descripteurs, des résumés propres à l'usager : c'est la réindexation des contenus par l'apport ou le vote des usages

**Mots clés :**

Recherche d'information (RI), réindexation, vote de l'usager, filtrage de contenus, analyse automatique, catégorisation documentaire.

**Abstract:**
Information retrieval (IR) is a user approach to obtain relevant information which meets needs with the help of a IR system (IRS). However, the IRS shows certain differences between user relevance and system relevance. These gaps are essentially related to the imperfection of the indexing process (as approach related to the IR), to problems related to the misunderstanding of the natural language and the non correspondence between the real needs of the user and the results of his query. As idea is to think about an "intellectual" indexing that takes into account the point of view of the user. By consulting the document, user can build information as added-value on the existing content: new information which grows contents and allows the semantic visibility or facilitates the reading by the annotations, by links to other content, by new descriptors, specific new abstracts of users: it is the reindexing of the contents by the contribution or the vote of the uses

**Keywords:**
Information retrieval (IR), reindexing, user vote, content filtering, automatic analysis, documentary categorization.

# Enrichissement des contenus par la réindexation des usagers :
*un état de l'art sur la problématique*

## 1. Introduction

Avec l'avènement du Web, la quantité d'informations disponible ne cesse pas de croître au cours de ces dernières années. Il a fallu donc envisager le développement des outils automatiques qui permettent l'accès ciblé et efficace à cette masse de données.

Plusieurs techniques sont concernées pour résoudre les problématiques liées à la recherche d'informations pertinentes.

En effet, la perpétuelle recherche d'optimisation donne naissance à des méthodologies de RI qui facilitent à l'utilisateur à la fois de déceler l'information et de formuler sa requête. Dans ce cadre, l'indexation est à la base du processus d'optimisation de méthodes de RI qui traitent l'information et participe à l'augmentation de sa valeur ajoutée.

Toutefois, l'indexation est un processus qui a montré ses limites sur la valeur informationnelle d'où l'idée d'enrichir ce processus par une nouvelle approche qu'est la réindexation par l'apport des usagers sur les contenus. Le principe est de valoriser l'implication des usagers dans leur lecture et l'importance de leur contribution dans l'enrichissement d'un contenu. Le but c'est d'augmenter les contenus par des nouvelles informations pertinentes et à valeur ajoutée. Ces informations peuvent être des commentaires qui permettent d'enrichir la sémantique d'un contenu, des liens vers d'autres contenus, des résumés propres à l'usager, des indications, des conseils, des annotations, des votes, des remarques, des évaluations, etc. [SIDHOM & DAVID, 2006] et c'est ainsi, qu'on arrivera à trouver de la valeur informationnelle autour des contenus. Dans cette considération, nous nous intéressons aussi aux contenus qui peuvent évoluer vers le multimédia. En exemple, les sources provenant des chaînes médiatiques telles que la radio et la télévision, et dans certains campus universitaires ; les supports pédagogiques accessibles dans les universités [FLORY, 2006] ; les communications d'entreprise sauvegardées en vidéoconférence au profit des praticiens ou universitaires, etc. Ces ressources peuvent être interconnectées au profit des besoins des usagers afin d'observer un réseau complet d'informations utiles à la communauté cible.

Notre article s'article sur les problématiques de l'enrichissement des contenus par les usagers dans une perspective de la réindexation. En section 2 et 3, nous présentons la problématique de la RI et les dimensions de recherche impliquées. Les sections 4 et 5 concernent l'identification de l'architecture d'un SRI et les processus inhérents avec les limites associées à l'indexation, à la formulation de la requête et à la fonction d'appariement entre requête et documents. Finalement, en section 6 avec la conclusion, la proposition d'une approche au problème pour s'orienter vers la réindexation par le vote des usagers et particulièrement les concepts de filtrage collaboratif et de la calculabilité du profil des usagers .

## 2. Recherche d'information

La recherche d'information est un domaine lié à toutes les sciences, mais historiquement lié aux sciences de l'information et de la bibliothéconomie. Dans ce domaine, le souci est d'établir des représentations sur les documents dans le but de représenter et de récupérer des connaissances des contenus documentaires à travers la construction d'index [BACHIMONT, 2004]. Les sciences informatiques par ses réflexions applicatives ont permis le développement d'outils pour traiter l'information et d'établir la représentation des documents au moment de leur indexation. Ainsi, le processus de recherche d'information dans sa complexité est rendu automatisable et donc rendu aisé dans sa manipulation par les utilisateurs. Intrinsèquement aujourd'hui, on peut dire que la recherche d'information est un champ transdisciplinaire, qui peut être étudié par plusieurs disciplines, par différentes communautés scientifiques et avec de multiples approches qui devraient permettre de trouver des solutions, d'améliorer l'efficacité du processus RI, de se rapprocher de l'indexation intellectuelle par des outils robustes [SIDHOM, 2002] et de ne plus s'attarder à intégrer les traitements sémantiques.

En effet le développement de l'Internet a permis de propulser la RI en avant scène avec de multiples applications tout en élargissant sa problématique par l'association de trois dimensions :

(i) les documents et leurs hétérogénéités : du texte au multimédia ;
(ii) les processus et leurs interopérabilités : de l'indexation au management des connaissances ;
(iii) les utilisateurs et leurs besoins informationnels : de l'indexeur, au veilleur et au filtrage collaboratifs pour la réindexation des contenus par les usagers.

Dans l'approche de développement d'outils, pour traiter l'information et pour établir la représentation des documents à l'usager, les informations sur les documents, sur la robustesse du processus d'indexation [SALTON, 1990] et sur le profil de l'utilisateur sont incontournables dans la réussite du processus RI. Sur l'aspect technique, quatre fonctions (ou modules) sont concurrents dans le processus, à savoir : -le stockage de l'information, -l'organisation de ces informations, -la recherche d'informations en réponse à des requêtes utilisateurs et -la restitution des informations pertinentes pour ces requêtes. A noter que seulement la dernière fonction reste visible pour l'utilisateur.

Après la présentation du champ transdisciplinaire de la RI, on s'oriente dans ce qui suit vers les problématiques liées à ce champ.

## 3. Problématique générale de la Recherche d'information

Le rôle d'un système de recherche d'information (SRI) est de permettre l'accès aux documents par leur contenu sémantique : l'usager est amené à indiquer le contenu qu'il souhaite retrouver dans les résultats de sa recherche.

Dans le processus général de la RI, nous distinguons trois niveaux d'interaction, à savoir :

### – *Le niveau utilisateur :*

A ce niveau, l'utilisateur exprime un besoin d'information sous forme d'une requête et il s'attend à obtenir des documents pertinents en réponse à ce besoin. La relation entre le besoin d'information et les documents attendus représente la *« relation de pertinence »*.

### – *Le niveau système :*

A ce niveau, le système RI répond à la requête formulée par l'utilisateur par un ensemble de documents stockés dans sa base de données. Souvent, la requête formulée par l'utilisateur est une description partielle de son vrai besoin d'information pour diverses causes : la limitation du langage de requêtes dans son pouvoir expressif et/ou l'ambiguïté inhérente à la requête et/ou sa traduction dans le langage [QUERO, 2007]. Egalement, les documents rendus par le SRI se limitent aux seuls contenus de la base et des documents répertoriés.

### – *Le niveau interne du système :*

La requête formulée par l'utilisateur, souvent exprimée en langage naturel, ne peut pas se comparer directement aux contenus des documents. Des représentations internes sont nécessaires pour rapprocher la sémantique de la requête à la sémantique des documents. Le processus de création de ces représentations est appelé « indexation ».

Dans la vision générale du processus RI, le développement des entreprises, des administrations et des institutions s'est fait dans des architectures réseaux de type Intranet ou Internet avec la circulation de leurs documents dans des opportunités de partage, de communication ou collaboration et d'action sur leur environnement.

Ce phénomène a entraîné l'essor de la problématique RI dans la dimension globale de la gestion électronique de documents pour permettre une circulation efficace de l'information, pour rapprocher les acteurs d'une organisation aux contenus ou encore pour capitaliser des connaissances partagées et vitales pour les acteurs de l'organisation.

A l'issue de la présentation des niveaux d'interaction dans un processus RI puis la dimension de la gestion électroniques de documents, on s'oriente dans ce qui suit vers l'architecture d'un SRI pour caractériser et expliquer les fonctions associées.

## 4. Architecture d'un système de recherche d'information

L'indexation est l'opération qui consiste à décrire et à caractériser un document à l'aide de représentations par les concepts de son contenu. Il est question, par un procédé intellectuel, automatique ou mixte, d'arriver à transcrire dans un langage documentaire (ie. artificiel) les concepts extraits des documents en passant par un processus d'analyse [QUERO, 2007]. Le processus d'analyse peut faire appel à des compétences intellectuelles du domaine (ie. l'indexeur humain), à la robustesse des outils (ie. des algorithmes complexes et des logiciels qui automatisent cette tâche d'indexation et se rapprochent de l'indexeur humain) ou à des compétences mixtes entre indexation automatique et réajustement par des compétences humaines. Le processus d'indexation constitue l'étape importante dans la valorisation d'un SRI.

Dans l'architecture générale d'un SRI, les phases de mise en oeuvre du processus relient des sous processus autonomes et communiquants : l'indexation des documents, la représentation sémantique des contenus (dans la requêtes ou les documents), l'appariement entre requête et documents, l'exploration des résultats de la recherche, etc. Dans La figure suivante (cf. figure1.), nous détaillons les principaux processus associés d'un SRI.

En référence à la norme AFNOR NF Z 47-102 1996, cette dernière donne une définition précise de l'indexation : *Cette étape consiste à analyser les documents et les requêtes afin de créer une représentation de leur contenu textuel qui soit exploitable par le SRI. Chaque document (et requête) est alors associé à un descripteur représenté par l'ensemble des termes d'indexation extraits.*

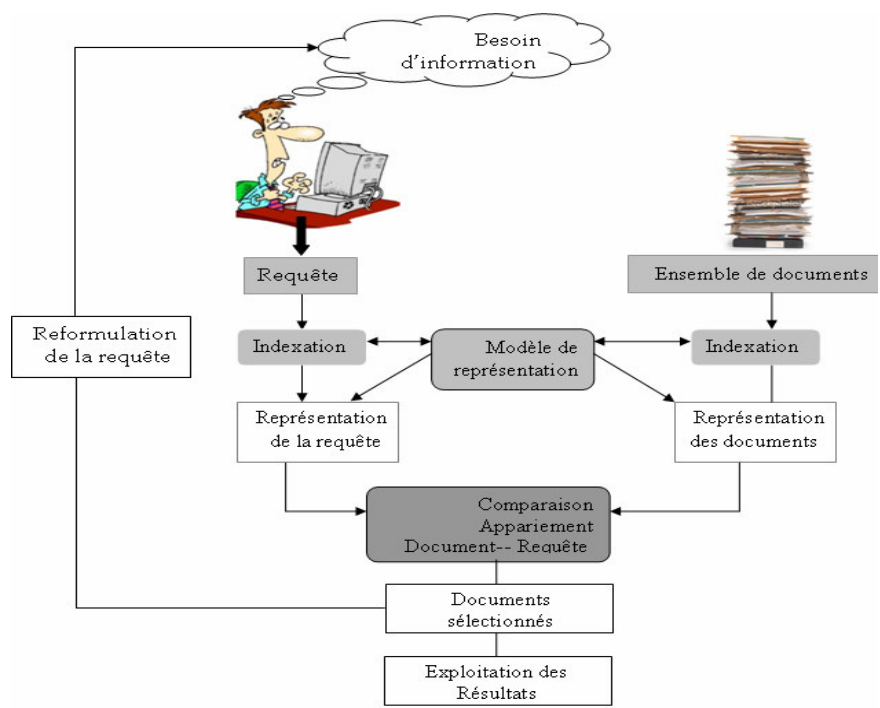

*Figure 1 : Processus en U de la recherche d'information.*

Dans les représentations internes du SRI, des fonctions associées au système sont d'une importance capitale et qui peuvent influencer considérablement son efficacité, à savoir :

– *Processus d'interrogation du système :*

Le processus d'interrogation agit entre la base d'index et la représentation de la requête de l'utilisateur. Elle permet d'effectuer la comparaison entre les deux représentations en utilisant une fonction d'appariement [JERIBI, 01].

– *Processus d'appariement requête– documents :*

Le processus d'appariement requête-documents permet de mesurer la pertinence des documents par rapport à une requête. En effet, à chaque saisie d'une requête, le système crée une représentation similaire à celle du document, puis il calcule un « score » de similarité (mathématique ou sémantique) qui traduit un degré de pertinence de la distance des termes de la requête avec ceux des documents. Le système retourne généralement des réponses ordonnées selon le score calculé.

Autre que les résultats du SRI et la pertinence calculée, l'utilisateur s'implique dans le dépouillement des documents réponses pour préserver ceux qui lui sont utiles. Son point de vue et ses appréciations sur les résultats du système seront utiles comme critères d'amélioration de ses performances, des processus associés et de la qualité calculée dans les réponses.

Néanmoins, dans l'architecture d'un SRI et quelques soient ses qualités, tout système ne peut être exploité de façon optimale et souvent des limites sont mesurées dans ses traitements. On voit ainsi que l'analyse des documents dans leur diversité est un domaine pluridisciplinaire qui implique des dimensions scientifiques connexes et des compétences complémentaires [ZACKLAD, 2004]. C'est ce qu'on compte expliquer dans ce qui suit par les limites des SRI.

## 5. Limites des systèmes de recherche d'information

L'information, qui était jusqu'aux dernières décennies, conservée sous une forme papier est désormais éclatée sur plusieurs supports informatiques et électroniques, dans plusieurs formats de document et dans des applications souvent isolées avec des objets multiples. De nos jours, une simple action requiert plusieurs opérations, comme répondre à un e-mail, cela nécessite une requête qui peut entraîner une cascade d'opérations consommatrices de temps, de processus qui requièrent d'intégrer l'information depuis une collection distribuée de supports : papier, formats électroniques, pages web, etc. Au final, il arrive qu'on ne parvienne pas à faire cette cascade d'opérations ou à trouver ce que l'on cherche, pour des raisons multiples : l'impatience, le manque de temps, le manque d'outils adéquats ou l'information est noyée dans un amas de documents peu visibles.

> *"Dans un cybermonde avec l'ergonomie et la robustesse des outils de recherche d'informations, toute personne devrait pouvoir trouver la bonne information, dans le bon format, avec le bon contexte et justement en réponse à son besoin formulé. Hélas, on ne vit pas dans un tel monde… si parfait…",*

Ainsi, le problème de la RI ne réside pas seulement dans la quantité des informations qui ne cesse de s'enrichir et donc de croître en volume de données mais plutôt dans l'aptitude cognitive humaine [JERIBI, 2001] ou à la robustesse des outils à retrouver l'information pertinente. Il est donc prévisible que le problème démontre des limites, à savoir :

– *Faiblesses liées à l'indexation*

En partie, à cause du non structuration de l'information et, en d'autre partie, la prolifération des formats de production documentaire ou informationnelle, que l'extraction de la connaissance significative et représentative du document est devenue une tâche difficile à gérer. On est confronté aux problèmes classiques liés aux langages naturels ou artificiels, et des traitements inhérents à chaque classe de langage (ie. polysémie, homonymie, anaphore, métaphore qui sont difficilement automatisables [LE LOARER, 1994]).

### – *Faiblesses liées à la formulation de la requête*

Pour formuler une requête, l'utilisateur fait appel à un langage naturel ou artificiel pour faire emprunt de mots clés souvent différents du vocabulaire employé par les auteurs des documents. Ainsi, l'appariement entre les deux classes de mots ou de langages, aussi bien les similarités cherchées ne seront pas si évidentes ou faciles par les processus d'analyse et de traitement. Certains problèmes peuvent parvenir à l'origine de :

- la requête qui ne tient pas en compte de la variation lexicale avec les contenus ;
- la disparité des variations morphologiques ;
- la disparité des variations syntaxiques et des constructions du langage ;
- la disparité des variations sémantiques et des constructions sémiques dans le langage par rapport aux objets de la recherche construits par l'utilisateur.

### – *Faiblesses liées à la fonction d'appariement document requête*

Souvent, la pertinence de l'appariement entre une requête et les documents repose essentiellement sur une comparaison entre les mots. Un document est jugé pertinent quand il recense tous les termes de la requête et/ou par association des termes sémantiquement proches de la requête. En exemple, un document qui identifie le terme d'indexation « voiture » ne pourra être apparié avec une requête représentée par le terme « automobile » si le système n'identifie pas le problème de la synonymie. Un pareil cas simple peut provoquer des insuffisances au niveau des performances du système en refusant de proposer à l'utilisateur des cas similaires qui sont susceptibles de l'intéresser.

En synthèse, les faiblesses ou limites liées aux SRI nous ont permis de clarifier des problèmes à résoudre dans notre travail tout en se proposant d'orienter nos réflexions de recherche vers la construction d'une plateforme de réindexation de contenus et de valoriser les aspects liés aux traitements de l'information et à la représentation des connaissances vis-à-vis de l'usager ou la dimension profil de l'utilisateur. Tel est notre propos dans la section suivante.

## 6. Proposition d'une approche : la réindexation par les usagers

Dans la chronologie générale de la documentation à l'hypertexte et de l'Internet au Web 2.0, les mutations ont constitué la partie la plus riche et la plus développée par l'aspect des outils. Cette chronologie entremêle plusieurs évolutions histoires, à savoir : (i) de la cybernétique et la documentation à l'information électronique, (ii) de l'hypertexte et l'Internet à la naissance du Web 2.0, le Web social (folksonomie) et les réseaux sociaux, et (iii) de la mise en place d'ARPANET et le développement de l'IST (ie. information scientifique et technique) à l'interconnexion progressive des réseaux d'information, d'archives ouvertes, etc.

Sur l'aspect (i), l'effet des technologies numériques a affecté de manière profonde les mutations sur le champ documentaire, qu'il s'agisse des procédés de travail intellectuel d'indexation, des pratiques et des usages de la recherche d'informations, des produits informationnels et ceux de la veille, etc. Ces mutations peuvent être appréhendées à partir des outils d'information et de leurs usages, qui contribuent à façonner ce champ. Le suivi et l'observation des incessantes évolutions techniques des outils de recherche sur Internet représentent ainsi une activité essentielle de veille. Les activités pédagogiques de formation de l'Information-Documentation, de la veille-IE et informatique doivent conduire aux évolutions de la fonction documentaire, de la fonction stratégique des données, de la fonction informations et connaissances, et plus particulièrement à la fonction du filtrage collaboratif et la calculabilité du profil des usagers dans le contexte du Web intelligent [CASTAGNOS & BOYER, 2004. Ainsi, La question sur les enjeux, notamment la "*maîtrise de l'information*" sur l'utilisateur et au profit de ses besoins, représente actuellement notre principal réflexion dans notre activité de recherche.

Sur l'aspect (ii), l'attention se tourne aux mutations actuelles des outils RI et des systèmes d'information vers les communautés virtuelles sur le Web. Ce qui représente un premier champ d'investigation dans la modélisation de l'utilisateur. C'est comme pour la notion d'hypertexte qui a montré la longue généalogie de ce principe d'organisation non-linéaire de l'information avec les techniques algorithmiques en informatique (ie. les pointeurs, les listes, les arbres et les graphes). L'émergence d'ARPANET, qui cherchait l'imbrication des réseaux informatiques, a accéléré des outils et des réseaux d'information à sortir des cloisonnements traditionnels qui enferment les outils RI dans une histoire purement "technique". Ainsi, nous sommes amenés à penser aux facteurs hétérogènes humains (ie. les réseaux sociaux, idéologiques, techniques ou culturels, les profils d'utilisateurs et les communautés d'intérêt sur le Web, etc.) dans un processus RI, dont dépendent les produits ou les services sur le Web. Cette réflexion s'inscrit aussi dans l'axe de notre recherche.

En dernier, sur l'aspect (iii), le cadre de la RI reste essentiel à améliorer les approches et les techniques, en ayant pour objectif principal la proposition d'une réindexation sur les contenus en prenant en compte le point de vue de l'usager. Le but est de valoriser des actions de l'usager qui consulte des documents ou produit d'information en laissant des marques de sa lecture afin d'améliorer le processus d'indexation et la sémantique des termes utilisés.

Dans ce contexte, notre proposition devrait permettre, par la souplesse de l'outil, de gérer l'information et son évolution vers le multimédia selon des procédés de même type que ceux mis en place pour l'information écrite (ie. indexation automatique, recherche par le contenu, extraction de connaissances, catégorisation des classes d'information et visualisation) [LATIRI, 2004]. Dès lors, les difficultés qui peuvent se manifester sont, d'une part, issues de la problématique liée à l'information multimédia (sur le web ou d'autres plates-formes) qui reste noyée dans d'autres contenus et, d'autre part, des manques d'informations sur le profil de l'usager, de sa modélisation ou de la catégorisation de ses traits cognitifs afin de capter ses préférences et de répondre aux mieux à ses besoins informationnels.

En outre, des problématiques liés aux SRI qui nous ont permis de définir des objectifs, la RI sur le multimédia reste à valoriser pour prendre en compte la sémantique véhiculée de son contenu sur le web et relativement aux domaines d'intérêt. Egalement, la question d'appréhender une nouvelle approche d'indexation par l'apport de l'usager reste prioritaire. Actuellement, les produits de la documentation et de l'indexation "intellectuelle" sur les contenus, sont en évolution de -l'approche d'indexation "fermée" ou le document est traité puis analysé par une communauté de spécialistes afin de construire un document secondaire qui valorise selon une méthodologie précise (ou grilles d'analyse) des informations descriptives et analytiques ; vers -l'approche d'indexation "ouverte" ou le document est valorisé par la prise en compte du vote de l'usager, ce dernier en consultant le document peut construire des informations à valeur ajoutée ou d'associer sont profil et ses traits d'intérêts sur le contenu.

Il s'agit de nouvelles informations qui enrichissent le contenu tout en permettant sa visibilité sémantique. Les annotations, les liens vers d'autres contenus, l'intégration de nouveaux descripteurs ainsi que des résumés propres à l'usager peuvent construire ce type de nouveaux contenus [SIDHOM, 2008].

# 7. Conclusion et perspectives

Dans cet état de l'art, qui est connexe aux problématiques d'indexation, de représentation des connaissances sur les contenus et des usagers de l'information, notre travail clarifie une nouvelle dimension dans l'approche de recherche d'informations dans le but d'intégrer le multimédia dans la gestion des contenus par l'apport des usagers ou leur vote : la réindexation par les usagers. A l'issue de cette réflexion, le processus de RI multimédia s'avère plus délicat que la RI textuelle par la nature du média et par la complexité du besoin des usagers sur le Web.

La contribution de l'usager sur le multimédia constituera l'originalité de notre problématique. Cet usager pourra s'impliquer à annoter et à relever des informations pertinentes selon sa culture, son profil, ses points de vue et ainsi mettre en évidence une sémantique sur le contenu grâce aux traces de sa lecture.

Ainsi, pour faire de la diffusion ciblée en fonction des besoins de l'usager ou du filtrage d'informations par rapport au profil de l'usager, il convient d'apprendre l'utilisateur courant, en modélisant ses besoins et ses préférences intrinsèques à partir des observations faites sur ses requêtes, ses navigations dans le contenu, ses partages d'informations avec d'autres, etc. *« Ces critères peuvent être considérés d'un point de vue quantitatif : il est possible d'évaluer numériquement l'utilité, la satisfaction, etc. On peut alors introduire la notion de pertinence : un item suggéré est d'autant plus pertinent qu'il répond à ces impératifs. »* [CASTAGNOS, 2008].

Ce processus « ouvert » aux communautés Web pour la valorisation de contenues constituera une nouvelle approche d'indexation : la réindexation par le vote des usages.